\documentclass{optica-article}
\journal{opticajournal} 

\articletype{Research Article}

\usepackage[utf8]{inputenc}
\usepackage[T1]{fontenc}
\usepackage{lineno}
\usepackage{xspace} 
\usepackage[normalem]{ulem}
\usepackage{soul}

\DeclareMathOperator*{\argmin}{argmin} 

\newfont{\cyr}{wncyr10 scaled 1200}%
\newcommand{\shah}{{\mbox{\cyr sh}}}%

\newcommand{\shuffle}{{\shah}}

\NewDocumentCommand{\revision}{m}{#1}
\NewDocumentCommand{\suppress}{m}{} 

\graphicspath{{./}{./images/}}

\makeatletter
\def\@journalname{Accepted for publication in Optics Express (https://doi.org/10.1364/OE.581902)}
\makeatother
\begin{document}
\title{Joint optical-digital design strategy for adaptive optics systems: application to wavelength selection for satellite imaging.}

\author{Florian Cheyssial,\authormark{1,*} Laurent M.\ Mugnier,\authormark{1} and Cyril Petit\authormark{1}}

\address{\authormark{1}DOTA, ONERA, Université Paris Saclay, BP 72, 92322 Châtillon cedex, France}

\email{\authormark{*}florian.cheyssial@onera.fr} 

\begin{abstract*} 
Adaptive optics can be used to mitigate the effects of atmospheric turbulence on imaging systems, but the correction is only partial and deconvolution is often required to improve the resolution. This results in entire optical/digital systems, which are traditionally designed sequentially, \textit{i.e.}, the adaptive optics system is optimised first, and the restoration algorithms are designed in a second time. Studies on optical/digital systems have shown that jointly optimising the whole system is a better alternative. We propose to extend these co-design strategies to the design of an adaptive optics assisted imaging system. We derive a simple criterion that takes into account the source properties and the entire optical/ digital system performance. To illustrate its interest, we use it to optimise the wavelength distribution between the wavefront sensor and the imaging camera. In addition, we explore the potential of using multiple imaging channels operating at different wavelength as a means of making an imaging system robust to turbulence strength and source magnitude variations. Later, any parameter of the optical/digital system, if not the entire system itself, could be optimised this way.
\end{abstract*}

\section{Introduction}

The increasing number of satellites in orbit, and more generally of space activities, has led to a growing interest in ground-based satellite imaging. The ability to produce high-resolution images of low Earth orbit (LEO) satellites using optical telescopes opens up a number of possibilities. For instance, high-resolution imaging can be used to monitor satellites throughout their lifecycle, from launch to deorbit, by controlling their condition and associated collision risks. It can also be used to detect and characterise poorly characterised or unknown objects, such as debris or asteroids. Imaging LEO satellites from the ground is challenging due to the atmospheric turbulence which distorts the wavefronts incident on the telescope. This can be compensated by the use of adaptive optics (AO) systems, which dramatically increase the resolution. However, this wavefront correction remains partial, and image restoration is therefore often required~\cite{conanEtudeCorrectionPartielle1994, conanImageFormationAdaptive1994, mugnierMyopicDeconvolutionWavefront2001}. A key parameter affecting the quality of the AO-corrected images is the imaging wavelength. For astronomical observations, this is usually fixed by the object of interest. But for satellite imaging, assuming we are not looking for a specific spectral signature, the imaging wavelength is a degree of freedom that can be optimised to produce the highest possible resolution images after deconvolution. This comes to be a particularly burning issue as the Onera is currently developing the future French system Providence dedicated to the observation of satellites, debris and small bodies of the solar system~\cite{petitPROVIDENCE25mAdaptive2024}.

The optimisation of the imaging wavelength for AO-assisted imaging systems was first studied by Tyler \& Fender who derived an analytical expression for the optimal imaging wavelength based on the Strehl ratio~\cite{tylerOptimalWavelengthSelection1994}. This expression has later been taken up by Rao \& Jiang who introduced a more precise expression of the Strehl ratio, especially in cases of partial AO correction~\cite{raoOptimumImagingObservation2002}. In both studies, the authors used a Strehl-based criterion to optimise the imaging wavelength, which is consistent with the way AO-assisted imaging systems are traditionally designed, using point spread function (PSF)-based criteria such as the residual wavefront error, the Strehl ratio or the PSF peak intensity~\cite{harmoni, roussetDesignNasmythAdaptive1998}. However, these criteria are only representative of the optical system quality, whereas the quantity of interest is usually the restored image quality. In particular, they neither take into account the noise in the acquisition process by the sensor nor the image deconvolution process. Consequently, optimising the imaging system, and \textit{a fortiori} the imaging wavelength, using such PSF-based criteria may lead to a sub-optimal design for the entire optical/digital system.

To avoid this problem, one solution is to jointly optimise the image acquisition system and the restoration algorithms using criteria based on the quality of the restored images. This approach is referred to as "joint optical/digital design" or "optical/digital co-design" in the literature~\cite{fontbonneComparisonMethodsEndtoend2022,storkInformationbasedMethodsOptics2006,storkTheoreticalFoundationsJoint2008}. The first example of a co-design approach was for the design of a phase mask to improve the depth of field of optical systems, by Cathey \& Dowski~\cite{catheyNewParadigmImaging2002}, but their method does not explicitly take into account the restoration process.
Stork and Robinson proposed a more general co-design methodology based on minimising the mean square error (MSE) between the true scene and the restored images~\cite{storkInformationbasedMethodsOptics2006,storkTheoreticalFoundationsJoint2008}. In particular, this approach can take into account the statistical properties of the object which is especially relevant in the case of satellite imaging where all images have a similar structure (a bright compact object on a dark background). 

In this paper, we optimise the imaging wavelength of an AO-assisted imaging system for satellite imaging using the framework proposed by Stork \& Robinson. A comparison with the results obtained using a PSF-based criterion is made and image simulations are performed to illustrate the visual gain. Because the optimal wavelength we obtain strongly depends on variable parameters, namely the seeing and the object magnitude, we also investigate the optimisation of the wavelength on average. In particular, we optimise the wavelengths of a multi-spectral imaging system in view of maintaining a satisfactory performance over a wide range of seeings and object magnitudes. All the imaging wavelength optimisations and other simulations presented afterwards are performed considering a PROVIDENCE-like system, \textit{i.e.}, a 2.5m AO-assisted optical telescope, which is described later. It is important to note that the PROVIDENCE design is currently under development and that the system parameters we use may not be representative of the actual system. Nonetheless, this is unimportant as the purpose of this article is to discuss the interest in using a co-design methodology for AO-assisted imaging systems rather than the results of the optimisations themselves.

The paper is organised as follows. We first discuss the wavelength optimisation issue in Sect.~\ref{sec:FD_issue}. The methodology we use is presented in Sect.~\ref{sec:model}, and we describe our simulation set-up as well as the different parameters that are used in Sect.~\ref{sec:setup}. In Sect.~\ref{sec:res_1chan}, we present the results of the optimisation of the imaging wavelength for a system possessing a single imaging channel. In Sect.~\ref{sec:disc_asp}, we discuss the choice of the optimisation metric and the assumptions assumed by our method. Finally, we present an extension of our framework in the case of a multi-channel imaging system, as well as the corresponding results, in Sect.~\ref{sec:ms}. We conclude on this work in Sect.~\ref{sec:ccl}.


\section{Flux distribution between the wavefront sensor and the scientific camera.}\label{sec:FD_issue}

In a telescope equipped with an AO system, the light collected by the entrance pupil is divided into at least two channels: the scientific channel which acquires the scientific images, and the wavefront sensor (WFS) channel which measures the wavefront distortions to feed the AO correction system. Optimising the imaging wavelength is therefore part of a wider issue, namely the distribution of the flux between these two channels. 

This issue of flux distribution comprises two separate questions. The first one is the photon distribution, \textit{i.e.}, how many photons are allocated to each channel. In the WFS channel, each photon helps measuring the turbulent wavefront. Hence, allocating more photons to the WFS channel improves the correction of the turbulence effects and increases the level of the optical transfer function (OTF). Conversely, allocating more photons to the imaging channel improves the image signal-to-noise ratio (SNR), resulting in a better deconvolution.

\begin{figure}[ht!]
    \centering
    \includegraphics[width=\textwidth]{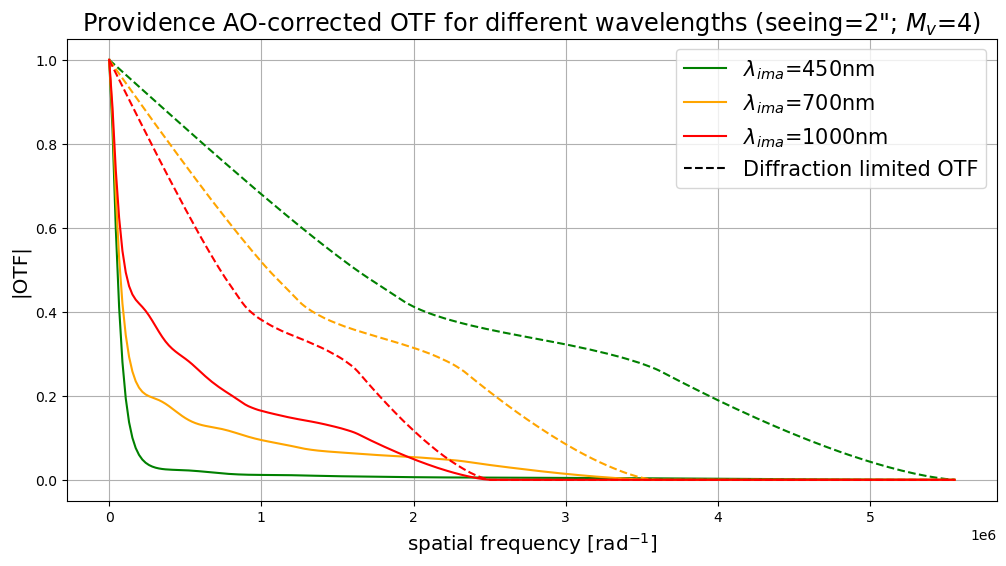}
    \caption{Simulation of the AO-corrected (filled lines) \& diffraction limited (dashed lines) OTFs modulus of the PROVIDENCE-like system described in Sect.~\ref{sec:setup}, for different imaging wavelengths, assuming a seeing of 2". The simulation details are described in Sect.~\ref{sec:setup}.}
    \label{fig:otf}
\end{figure}

The second question is the wavelength distribution, \textit{i.e.}, which photons are allocated to each channel. For the scientific channel, the photons' wavelengths greatly influence the image resolution. Two competing effects occur when the imaging wavelength gets shorter: \textit{i)} diffraction effects decrease, leading to a sharper PSF and, equivalently, a higher cut-off frequency of the OTF at the diffraction limit (see the dashed lines in Fig.~\ref{fig:otf}) ; \textit{ii)} the residual phase error increases, causing the corrected PSF to deviate from the diffraction-limited one and the OTF global level to drop (see full lines Fig.~\ref{fig:otf}). 

Since PSF-based criteria do not take into account the image SNR, only the wavelength distribution issue has been addressed in the aforementioned previous works (\cite{tylerOptimalWavelengthSelection1994, raoOptimumImagingObservation2002}). However, the image SNR has a significant influence on both the focal plane image quality and the restoration process. Thus, even if deconvolution were not considered, the SNR should influence the optimal imaging wavelength value. Additionally, it can be observed on Fig.~\ref{fig:otf} that, depending on the spatial frequency considered, the wavelength that gives the highest OTF varies. Hence, the object's spatial spectrum should also influence the optimal imaging wavelength. These observations justify considering a co-design approach. 


\section{The co-design framework}\label{sec:model}

This section describes our methodology for optimising the flux distribution in an AO-assisted imaging system. It requires: a model of the optical system, and in particular of its OTF, which can be used to simulate images; a reconstruction method that is able to restore simulated images; and a metric that quantifies the quality of the restored images. It is worth mentioning that the following framework is not specific to wavelength selection, and can be used to optimise any parameter of the imaging system.


\subsection{Optimisation criterion}

The choice of the optimisation criterion is important and must be made in accordance with the desired objective, such as identifying a satellite or detecting a variation in its attitude. In this paper, we do not consider any particular case, and we simply define the objective as: producing restored images that are as close as possible to the observed objects. Besides, we want the chosen criterion to be representative of the system performance on average. To this end, we consider a criterion $R_D$ that takes the form of a mean distance (in the mathematical sense) between the possible objects $\textbf{o}$ and their estimates $\hat{\textbf{o}}(\textbf{i})$, which are obtained by the restoration algorithm from the images $\textbf{i}$:
\begin{equation}
    R_D(\hat{\textbf{o}})\ =\ \mathbb{E}_{\textbf{o},\textbf{i}}\left[D\left(\hat{\textbf{o}}(\textbf{i}),\textbf{o}\right)\right],
    \label{eq:mse}
\end{equation}
where $D$ is a distance function, and $\mathbb{E}_{\textbf{o},\textbf{i}}$ stands for the mathematical expectation over the possible objects and noise realisations. Among the possible distances for image quality assessment, we consider the square Euclidean norm, also known as the L2 norm. This gives for $R_D$ the well-known mean-square-error (MSE), which is one of the most used optimisation criteria. There are two main benefits to using the MSE. Firstly, thanks to Parseval's theorem, the MSE can be expressed in either the direct or Fourier domain. This enables a frequency analysis of the error and therefore a better understanding of the results. Secondly, if the linear minimum mean-squared-error (LMMSE) estimator is used to restore the images, the MSE can be computed using an analytical formula, as presented in Sect.~\ref{sec:reconstruction}. This drastically reduces the computational burden of the optimisation, as it is no longer necessary to simulate numerous restored images to compute the average error.\par

Two other widely used distances for image quality assessment are the L1 norm and the structural similarity index measurement (SSIM)~\cite{wangImageQualityAssessment2004}. Although their corresponding criteria do not have the same advantages as the MSE, these distances may, depending on the type of images, be more representative of the image quality~\cite{gomesAssessingQualityRestored2017, wangMeanSquaredError2009}. A comparison of the results obtained using these distances or the L2 norm will be presented in Sect.~\ref{sec:ometrics}. 

\subsection{Imaging model}

The focal plane image $\textbf{i}$ of a continuous object $o$ is described by the following relation:
\begin{equation}
    \textbf{i}\ =\ [h_{det}\ast h_{opt}\ast o]_\shuffle + \textbf{n},
\end{equation}
where $\textbf{n}$ is an additive noise, supposed independent from the object, $h_{det}$ is the detector PSF and $h_{opt}$ is the AO corrected optical PSF. Italic letters are used to represent continuous variables, while bold letters represent discrete value vectors. We assume that the image is taken within an isoplanetic patch, \textit{i.e.}, that the PSF doesn't vary with position. The symbols $\shuffle$ and $\ast$ stand respectively for the discretisation and convolution operators. This relation can be approximated by a discrete convolution,
\begin{equation}
    \textbf{i}\ =\ \textbf{h}\ast \textbf{o} + \textbf{n},
    \label{eq:img_model}
\end{equation}
where $\textbf{o}$ is the decomposition of $o$ on some discrete basis, \textit{e.g.}, cardinal sines, and $\textbf{h}$ is a discrete version of $h_{det}\ast h_{opt}$ (which depends on the basis for $\textbf{o}$). The AO corrected PSF $h_{opt}$ can be decomposed into a static PSF $h_{static}$, which includes the telescope static aberrations and the diffraction, and a turbulent PSF $h_{ao}$ that embodies the residual phase effects on the PSF~\cite{roddierEffectsAtmosphericTurbulence1981,conanEtudeCorrectionPartielle1994}:
\begin{equation}
    h_{opt} = h_{static}\ast h_{ao}.
    \label{eq:rodier}
\end{equation}


\subsection{Object prior and noise model}\label{sec:prior}

To model the possible objects, we consider a spatially homogeneous Gaussian prior defined by its mean $\overline{\textbf{o}}$ and power spectral density (PSD) $S_o$. The PSD of natural scenes, including satellite images, can be described by the following parametric model, which is a slightly modified writing of Matern’s model~\cite{yanMarginalizedMyopicDeconvolution2023, kattnigModelSecondorderStatistic1997,conanMyopicDeconvolutionAdaptive1998}:

\begin{equation}
    S_o(\mathrm{f})\ =\ \frac{A^2}{k+\mathrm{f}^p},
    \label{eq:psdo}
\end{equation}

\noindent
where $\mathrm{f}=|f|$ is the modulus of the two-dimensional spatial frequency $f$. In this model: $A$ gives the global PSD level, $p$ characterises the PSD decrease at high frequencies and $k$ sets the PSD level for very low spatial frequencies. We use the same model to describe the PSD of the discrete object. In practice, a working value for $A$ is the average object total flux in photoelectron $N_{ph}$. As for the decrease rate $p$, its value is around 2 to 2.5 for typical satellite images. Regarding the mean object $\overline{\textbf{o}}$, we simply define it as a constant whose value is the average flux per pixel $N_{ph}/N_{pix}$, where $N_{pix}$ is the number of pixels in the image.\par

The additive noise term $\textbf{n}$ may include readout noise and photon noise. Under high flux assumptions, it can be well approximated by an inhomogeneous zero-mean white Gaussian noise. In this paper, we further approximate the noise as homogeneous (\textit{i.e.}, with a constant variance) such that it has a PSD $S_n$, which drastically simplifies the expression of the MSE (see Sect.~\ref{sec:reconstruction}). The noise PSD value is set to $S_n=N_{ph} + \sigma^2_{ron}\times N_{pix}$, where $\sigma^2_{ron}$ is the readout noise (RON) variance, in photoelectron per pixel, of the imaging sensor and the $N_{ph}$ term embodies the photonic noise contribution. It must be noted that this approximation is somewhat crude since images of satellites mainly consist in a bright compact object on a dark background, causing the photonic noise to be locally distributed. The impact of this approximation on the results is discussed in more details in Sec\revision{t.}~\ref{sec:vs_noise}. 


\subsection{Restoration algorithm}\label{sec:reconstruction}

To restore the images, we use the minimum mean square error (MMSE) estimator as it minimises the MSE which is our optimisation criterion. Under the assumptions previously established on the object prior and noise distribution, the MMSE is equal to its linear form (LMMSE) and corresponds to a Wiener filter. In addition, if we make an approximation of periodicity for the object and the PSF, the MMSE estimate $\tilde{\hat{\textbf{o}}}_{mmse}$ can be written in the discrete Fourier domain:
\begin{equation}
    \tilde{\hat{\mathrm{o}}}_{mmse}(f_j) = \frac{S_o(f_j)\tilde{\mathrm{h}}^*(f_j)\tilde{\mathrm{i}}(f_j)+S_n(f_j){\tilde{\overline{\mathrm{o}}}}(f_j)}{S_o(f_j)\vert\tilde{\mathrm{h}}(f_j)\vert^2+S_n(f_j)},
    \label{eq:wiener}
\end{equation}
where the tilde denotes the two-dimensional discrete Fourier transform (DTF) and the discrete frequencies $f_j$, \revision{$\forall j\in\llbracket0,N_{pix}-1\rrbracket$}, are the Fourier conjugated variables of the image pixels position (in radians).

Using the previous equations, it is possible to show that the MSE (Eq.~\eqref{eq:mse}) between the true object $\textbf{o}$ and its estimation by MMSE (Eq.~\eqref{eq:wiener}) is given by the following analytical formula:
\begin{equation}
    \text{MSE}(\hat{\textbf{o}}_{mmse},\textbf{o})\ =\ \frac{1}{N_{pix}}\sum_{j=0}^{N_{pix}-1} \frac{S_n(f_j)}{\vert\tilde{\mathrm{h}}(f_j)\vert^2+\frac{S_n(f_j)}{S_o(f_j)}}.
    \label{eq:mse_wiener}
\end{equation}

In practice, we use a slightly different criterion, that we call root normalised MSE (RNMSE), which is defined as:
\begin{equation}
    \mathrm{RNMSE} = \sqrt{\frac{MSE}{N_{ph}^2}} 
    \label{eq:rnmse}
\end{equation}
Normalising the MSE by $N_{ph}^2$ allows us to compare the entire system performance for objects of different fluxes, and taking its root lead to a criterion with the same unity as the object. The RNMSE hence corresponds to a relative deviation between two images. If there is no difference then RNMSE=0, and if the restored object is 0 then RNMSE=1. What's really remarkable here is the fact that, using Eq.~\eqref{eq:mse_wiener}, the average performance of the entire optical/digital system over a given set of objects can be calculated without computing a single image.


\section{Simulation setup \& optimisation process}\label{sec:setup}

In order to find the optimal flux distribution, we need to plot the RNMSE as a function of this distribution. According to Eq.~\eqref{eq:mse_wiener} and Eq.~\eqref{eq:rnmse}, computing the RNMSE requires the knowledge of: the OTF $\tilde{\textbf{h}}$, the object PSD $S_o$ and the noise PSD $S_n$. To simulate the AO-corrected OTF, we use a Fourier-based AO simulation method. These methods, explained in detail in~\cite{jolissaintAnalyticalModelingAdaptive2006, fetickIncludingPyramidOptical2023}, use physics-based analytical formulae, which makes them much faster than end-to-end simulations. This acceleration is of particular interest to us as we need to perform numerous OTF simulations during the optimisation. Table~\ref{tab:params} summarises the key parameters of the AO system and telescope designs that are taken into account by our simulation tool. Note that, in particular, our simulation tool explicitly takes into account the flux distribution between the wavefront and the imaging channels, both in terms of number of photons and in terms of wavelength. The parameter values listed in Tab.~\ref{tab:params} correspond to those of the PROVIDENCE-like system for which we wish to optimise the imaging wavelength. As we are not considering the presence of a laser guide star here, the reference for the WFS measurements is the target satellite being imaged itself. The values in Tab.~\ref{tab:params} for the WFS reference correspond roughly to a 10m satellite orbiting at 800km and being observed at an elevation of 60°. Besides the parameters listed in Tab.~\ref{tab:params}, the simulated OTFs also depend on the turbulence conditions which is specified using a turbulence profile and the seeing to give the global turbulence strength. In all the following, the seeing is given at 500\,nm at the zenith.

\begin{table}[htbp]
\caption{Simulation fixed parameters (frame delay not including WFS integration time).}
  \label{tab:params}
  \centering
\begin{tabular}{ccccc}
\hline
 & Parameter & Notation & Value & Unit\\
\hline\hline
\multirow{3}{4em}{Telescope design} & Diameter & $D$ & 2.5 & [m] \\
& Central obstruction & $O_c$ & 0.75 & [m] \\
& Image pixel size & $\theta_{pix}$ & $8\times10^{-8}$ & [rad]\\ 
\hline
\multirow{8}{4em}{AO system design} & loop frequency & $f_{ao}$ & 2000 & [Hz] \\
& frame delay & $\varnothing$ & 0.5 & [ms] \\
& Number of sub-apertures & $N_{sa}$ & 19$\times$19 & $\varnothing$ \\
& Number of actuators & $N_{act}$ & 20$\times$20 & $\varnothing$ \\
& WFS readout noise & $\varnothing$ & 0.5 & [$e^-$/pix] \\
& WFS diffraction sampling & $\varnothing$ & Shannon/2 & $\varnothing$ \\ 
& WFS wavelength & $\lambda_{wfs}$ & 700 & [nm]\\
& WFS bandwidth & $\Delta\lambda_{wfs}$ & 400 & [nm]\\
\hline
\multirow{2}{4em}{Target} & Size & $\theta_{sat}$ & 2.25 & [arcsec] \\
& Elevation & $\epsilon$ & 60 & [°] \\
\hline
\end{tabular}
\end{table}

The PSDs $S_o$ and $S_n$ can be computed using the expressions given in Sect.~\ref{sec:prior}. These expressions depend on the hyperparameters $p$ and $k$, both of which are set to 2, as well as on the number of photoelectrons per image $N_{ph}$. The number of photoelectrons per frame and sub-aperture in the WFS, $N_{ph,wfs}$, is also required to compute the OTF. Both $N_{ph}$ and $N_{ph,wfs}$ can be derived from the spectral irradiance at the pupil plane of the telescope $E_{pp}(\lambda)$:
\begin{equation}
    \begin{split}
        N_{ph} &= \frac{\tau_{ima}S_{tel}}{\hbar c}\int M_{ima}(\lambda)E_{pp}(\lambda)T_{tel}(\lambda) Q_{ima}(\lambda)\lambda \mathrm{d}\lambda\\
        N_{ph,wfs} &= \frac{\tau_{wfs}S_{sa}}{\hbar c}\int M_{wfs}(\lambda)E_{pp}(\lambda)T_{tel}(\lambda) T_{wfs}(\lambda)Q_{wfs}(\lambda)\lambda \mathrm{d}\lambda\\
    \end{split}
\end{equation}
where $c$ is the speed of light in vacuum and $\hbar$ is the Planck constant. The integration times of the imaging sensor and of the WFS are respectively denoted by $\tau_{ima}$ and $\tau_{wfs}$. In the following, we neglect the readout time of the WFS and approximate $\tau_{wfs}$ to $1/f_{ao}$, where $f_{ao}$ is the AO loop frequency. The collecting surface of the telescope is denoted by $S_{tel}$, while $S_{sa}$ stands for the surface of a sub-aperture of the WFS projected onto the pupil plane. The different transmission terms are represented by the letter $T$, with $T_{tel}$ the full optical system transmission (from the entrance pupil to the imaging sensor), and $T_{wfs}$ a term that compensates for the transmission discrepancies between the imaging and WFS channels ($T_{wfs}$ can therefore exceed 1). The terms $Q_{ima}$ and $Q_{wfs}$ stand for the quantum efficiencies in their respective channels. Finally, $M_{ima}$ and $M_{wfs}$ are transmission functions whose values range from 0 to 1, defining the flux distribution across the two channels. These are the functions that we seek to optimise. To compute the irradiance in the telescope's pupil plane $E_{pp}$, we use the fact that satellites are secondary sources that reflect sunlight. Hence, an average value of the irradiance $E_{pp}$ can be derived from the solar spectral luminance and the magnitude of the satellite in the V band $M^v$. In the end, for each channel, our photometric model takes into account: the exposure time, the effective area in the telescope's pupil plane, the transmission of the optics, the quantum efficiency, the spectrum of the object and the transmission function of each channel  ($M_{ima}$ and $M_{wfs}$) that are to be optimised.

For simplicity, we will assume in this study that the imaging channel uses all the flux in a spectral band of central wavelength $\lambda_{ima}$ and width $\Delta\lambda_{ima}$. Then the WFS channel collects all the remaining flux in a larger band of central wavelength $\lambda_{wfs}$ and width $\Delta\lambda_{wfs}$, \textit{i.e.}:
\begin{equation}
    \left\{
    \begin{aligned}
        M_{ima}(\lambda) &= \sqcap\left(\frac{\lambda-\lambda_{ima}}{\Delta\lambda_{ima}} \right)\\
        M_{wfs}(\lambda) &= \sqcap\left(\frac{\lambda-\lambda_{wfs}}{\Delta\lambda_{wfs}} \right)\times\left( 1-M_{ima}(\lambda)\right)
    \end{aligned}\right.
    \quad ;\quad \sqcap (x) = 
    \left\{
    \begin{aligned}
        &1,\ if\ -0.5\le x\le0.5\\
        &0,\ elsewhere 
    \end{aligned}
    \right.
\end{equation}
In practice, we chose to fix the values of $\lambda_{wfs}$ and $\Delta\lambda_{wfs}$, such that only $\lambda_{ima}$ and $\Delta\lambda_{ima}$ remain to be optimised.

Ultimately, for given values of seeing and $M^v$, the optimisation process is as follows: \textit{i)} we modify the values of $\lambda_{ima}$ and $\Delta\lambda_{ima}$; \textit{ii)} we compute the corresponding distribution functions $M$ and the corresponding number of photoelectrons $N_{ph}$; \textit{iii)} we simulate the FTO and compute the PSDs $S_o$ and $S_n$; \textit{iv)} we compute the resulting MSE. 


\section{Optimisation of the imaging channel wavelength and bandwidth}\label{sec:res_1chan}
\subsection{Optimisation of the imaging central wavelength}\label{sec:lbd_ima}

\begin{figure}[ht!]
    \centering
    \includegraphics[scale=0.58]{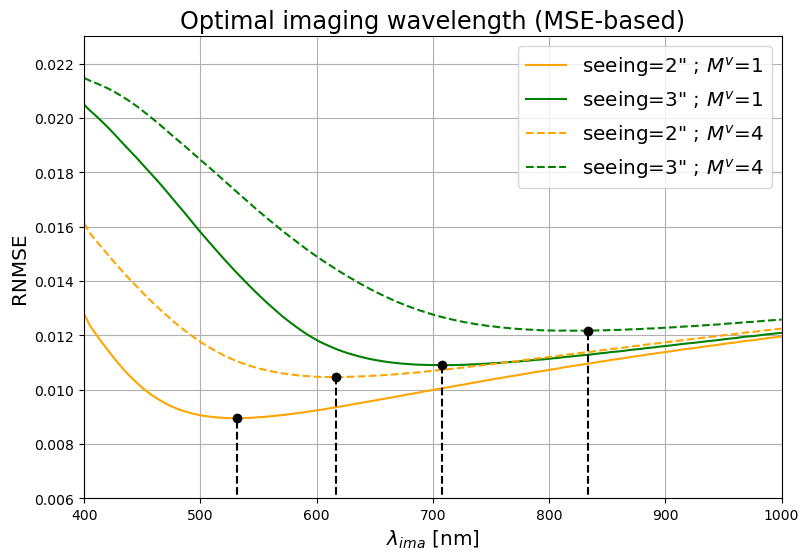}
    \caption{Root normalised MSE error depending on $\lambda_{ima}$ for different seeings and object magnitude ($\Delta\lambda=50$\,nm). The black dots indicate the minimum of each curve.}
    \label{fig:lbd_ima}
\end{figure}

To start with, we study the influence of the imaging central wavelength $\lambda_{ima}$ alone. We assume it can vary from $400$\,nm to $1000$\,nm, and we fix the imaging channel bandwidth $\Delta\lambda_{ima}$ to $50$\,nm. This value is wide enough to have a non negligible flux in the imaging channel, but still small enough for the PSF to be approximated as the one at $\lambda_{ima}$. The values of $\lambda_{wfs}$ and $\Delta\lambda_{wfs}$ are respectively fixed to 700\,nm and 400\,nm, \textit{i.e.}, the WFS collects the entire flux within the range [500\,nm,\,900\,nm] minus the imaging band. We plot in Fig.~\ref{fig:lbd_ima} the RNMSE as a function of the imaging central wavelength for two seeing conditions and two object magnitudes. The black dots mark the optimal imaging wavelength for each case. For wavelengths greater than the optimum, we can see that the error increases because diffraction increasingly limits the system's resolution. For shorter wavelengths, the error increases due to the residual phase error. Hence, for given conditions (seeing and magnitude), the optimal imaging wavelength corresponds to the best compromise between diffraction and residual phase limitations.

We can observe that as the seeing increases (\textit{i.e.}, as the turbulence gets stronger) for a given object magnitude, the error increases and the optimal wavelength becomes longer (compare the yellow and green curves). This is because the residual phase error increases with the seeing, whereas the diffraction effects remain constant, hence the optimum is pushed towards longer wavelengths. Similarly, as the magnitude increases (\textit{i.e.}, as the object becomes fainter) for a given seeing, we can observe that both the RNMSE and the optimal imaging wavelength increase (compare the solid and dashed curves). 

The dependency of the optimal imaging wavelength on the turbulence strength can be observed using PSF-based criteria and has already been mentioned in previous works \suppress{( }\cite{raoOptimumImagingObservation2002, tylerOptimalWavelengthSelection1994}\suppress{)}. These criteria are however insensitive to the image SNR, and the dependency of the optimal imaging wavelength on the source magnitude is, to our knowledge, a new finding. For comparison, we applied the criterion proposed by Rao \& Jiang to our simulated PSFs, and it resulted in significantly longer optimal imaging wavelengths. For example, when the seeing is $2"$, the optimal wavelength thus obtained is 919\,nm, which is far from the 505\,nm (when $M^v=0$) or 616\,nm (when $M^v=4$) we get using the MSE.

\subsection{RNMSE spectrum}\label{sec:rmse_spectrum}

To go further into the analysis of the curves presented in Fig.~\ref{fig:lbd_ima}, and in particular to understand the evolution of the RNMSE with the object magnitude, we can look at the normalised MSE spectral density $\varepsilon$, defined as:
\begin{equation}
    \varepsilon(f_j) = \frac{S_n(f_j)}{N_{ph}^2N_{pix}\left(\vert\tilde{\mathrm{h}}(f_j)\vert^2+\frac{S_n(f_j)}{S_o(f_j)}\right)}\qquad ;\qquad s.t.:\ \mathrm{RNMSE}=\sqrt{\sum_{j=0}^{N_{pix}-1}\varepsilon(f_j)}.
\end{equation}

\begin{figure}[ht!]
    \centering
    \includegraphics[scale=0.5]{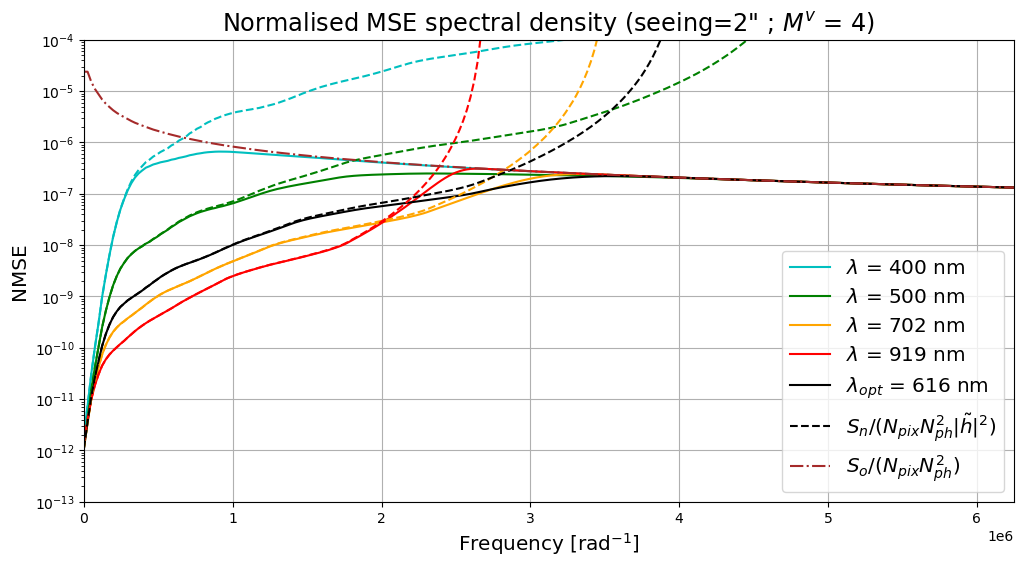}
    \caption{NMSE spectrum for different imaging wavelengths $\lambda_{ima}$ (seeing=2"; $M^v$=4; $\Delta\lambda=50$\,nm).}
    \label{fig:mse_freq}
\end{figure}

In Fig.~\ref{fig:mse_freq}, we plot the circular sum of $\varepsilon$ as a function of the radial frequency $\mathrm{f}$ for different imaging wavelengths (solid coloured lines), a seeing of 2" and $M^v=4$. We also plot in this figure the asymptotic curves of $\varepsilon$ when the image SNR - defined as $S_o|\tilde{\mathrm{h}}|^2/S_n$ - tends to either zero or infinity. When the image SNR tends to infinity, the Wiener filter behaves as an inverse filter (\textit{i.e.}, $\tilde{\hat{\textbf{o}}}_{mmse}=\tilde{\textbf{i}}/\tilde{\textbf{h}}$), and $\varepsilon$ tends to $S_n/(N_{ph}^2N_{pix}|\tilde{\mathrm{h}}|^2)$ (dashed coloured lines) which correspond to the noise amplification by inverse filtering. This is the ideal situation. Conversely, when the image SNR tends to 0, the Wiener filter returns the mean object $\overline{\textbf{o}}$ (\textit{i.e.}, there is no deconvolution) and $\varepsilon$ tends to $S_o/(N_{ph}^2N_{pix})$ (dashed dotted line). 

We can observe on Fig.~\ref{fig:mse_freq} that, for all wavelengths, $\varepsilon$ is first tangent to the infinite image SNR asymptotic curve at low frequencies. In other words, the object's low spatial frequencies are reconstructed through inverse filtering. Then, as the OTF decreases (with the frequency), the solid and dashed lines start separating. This corresponds to the intermediate regime where the deconvolution is regularised by the object prior. Finally, when the OTF gets too small, $\varepsilon$ smoothly converges toward $S_o/(N_{ph}^2N_{pix})$, which marks the point from which the object's frequencies are no more restored. For the longest imaging wavelengths (see, \textit{e.g.}, the red and yellow curves), this transition from one regime to an other is rapid and happens just before the cut-off frequency of the system since, for these wavelengths, the OTF remains pretty high until the cut-off (see the red curve on Fig.~\ref{fig:otf}). At the opposite for the shortest imaging wavelengths, the OTF quickly converges to zero (see the green curve in Fig.~\ref{fig:otf}), and we can see that it causes $\varepsilon$ to rapidly join $S_o/(N_{ph}^2N_{pix})$ (see the blue curve). For intermediate wavelengths, the transition is smoother and $\varepsilon$ matches $S_o/(N_{ph}^2N_{pix})$ at greater frequency. 

The optimal imaging wavelength corresponds to the wavelength for which the integral of $\varepsilon$ is the smallest (here the black curve). Therefore, this wavelength does not necessary minimises $\varepsilon$ at all frequencies. In particular we can see that at low frequencies, the reconstruction error decreases as the imaging wavelength increases. Indeed, at low frequencies, the OTF increases with the wavelength (see Fig.~\ref{fig:otf}). In practice, we noticed that the optimal imaging wavelength is also the one -~or at least very close to the one~- for which the intersection between the two asymptotic curves is at the highest frequency. This observation helps understanding the evolution of the RNMSE with the seeing or the object's magnitude. When the seeing increases, neither $S_o$ nor $S_n$ vary, but the OTF decreases, which makes all the solid and dashed lines rise while the dashed dotted line remains the same. As a result, $\varepsilon$ -~and therefore the RNMSE~- increases, and the wavelength for which the two asymptotic curves crosses each other the furthest increases as well. Similarly, when the magnitude increases (\textit{i.e.}, when $N_{ph}$ decreases), the OTF barely vary, $S_o/N_{ph}^2$ remains constant, and $S_n/N_{ph}^2$ increases. Consequently, the solid and dashed lines rise while the dashed dotted line remains the same, just like when the seeing increases. 

\subsection{Illustrations}

\begin{figure}[ht!]
    \centering
    \includegraphics[width=\textwidth]{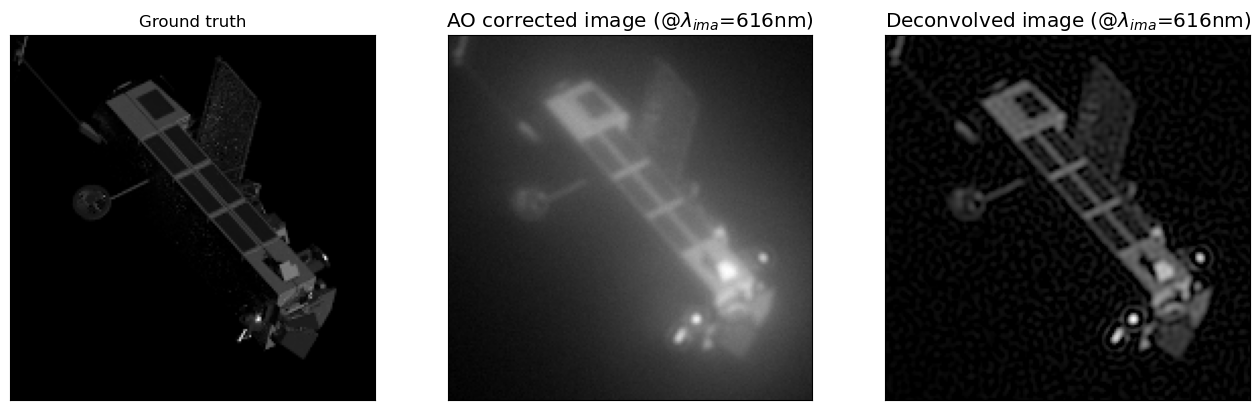}
    \caption{Simulation of an observation of the satellite Envisat at $\lambda_{ima}$=616\,nm (seeing=2"; $M^v$=4). \textit{Left}: True object; \textit{middle}: AO-corrected focal plane image; \textit{right}: Wiener-restored image.}
    \label{fig:sats0}
\end{figure}

\begin{figure}[ht!]
    \centering
    \includegraphics[width=\textwidth]{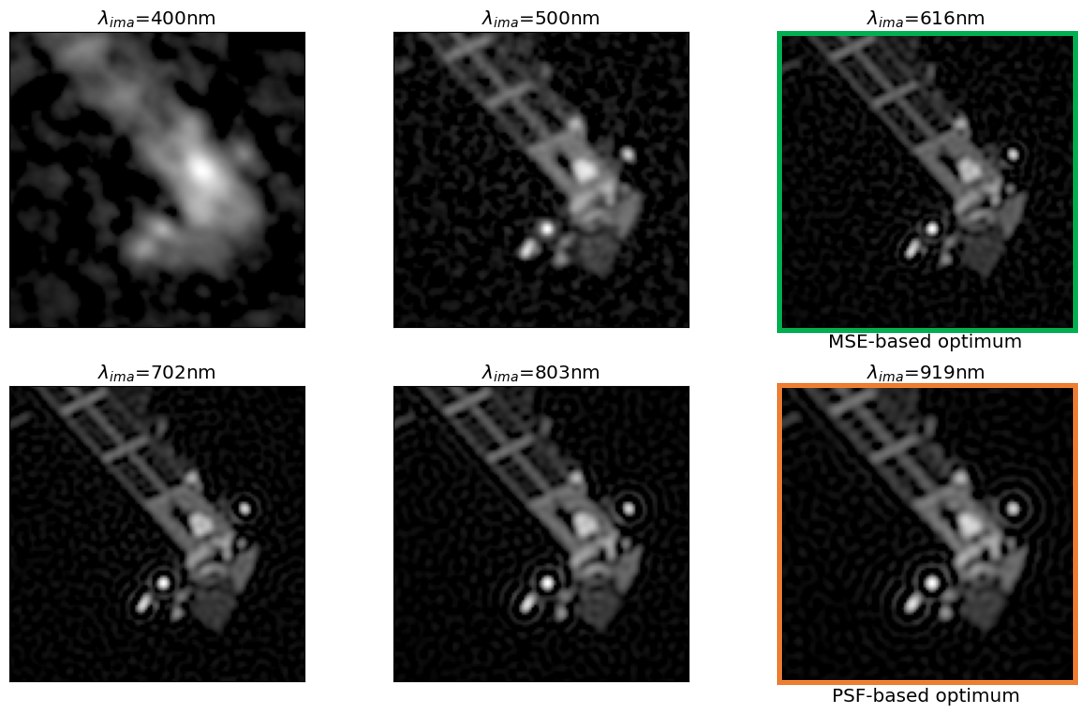}
    \caption{Simulated images, after deconvolution, of observations of the satellite Envisat at different wavelengths (seeing=2"; $M^v$=4). The images are zoomed in.}
    \label{fig:sats}
\end{figure}

In order to verify that these RNMSE curves are indeed consistent with the image quality at the different wavelengths, we simulate AO-corrected focal plane images of the satellite Envisat using Eq.~\eqref{eq:img_model}, and we deconvolve them with the Wiener filter of Eq.~\eqref{eq:wiener}. An example of a focal plane image and its deconvolution are given in Fig.~\ref{fig:sats0}, with the reference object used for the simulations on the left of the figure.
In Fig.~\ref{fig:sats}, we show some restored images at different wavelengths, including the MSE-based optimum (see the green surrounded image) and the PSF-based optimum (orange surrounded image), for the case where the seeing is 2" and $M^v=4$. We can first see that, except for the image at 400\,nm which has by far the highest error, all the images are of pretty good quality. In particular, the images at 616\,nm (the optimum) and 700\,nm, which are almost indistinguishable, appear to be "the best ones", in the sense that they are the most resolved and the closest, visually, to the ground truth. This is consistent with the results of Fig.~\ref{fig:lbd_ima} (yellow dashed curve) where we can see that: \textit{i)} the RNMSE values are close for these two wavelengths, \textit{ii)} these values correspond to the minimum of the curve. Similarly, the images at 500\,nm and 919\,nm also have the same quadratic error w.r.t.\ the ground true, but the images are quite different. The image at 500\,nm exhibits more reconstruction artefacts than the one at 919\,nm, but we can see by looking at the satellite's details and edges that it is also more resolved. This is compatible with the error spectrums presented in Fig.~\ref{fig:mse_freq}. Indeed, at 919\,nm, the low spatial frequencies are very well reconstructed and the error is essentially composed of mid/high frequencies, while at 500\,nm, the error is more distributed.   



\subsection{Joint optimisation of the imaging wavelength and bandwidth}

\begin{figure}[ht!]
    \centering
    \includegraphics[scale=0.58]{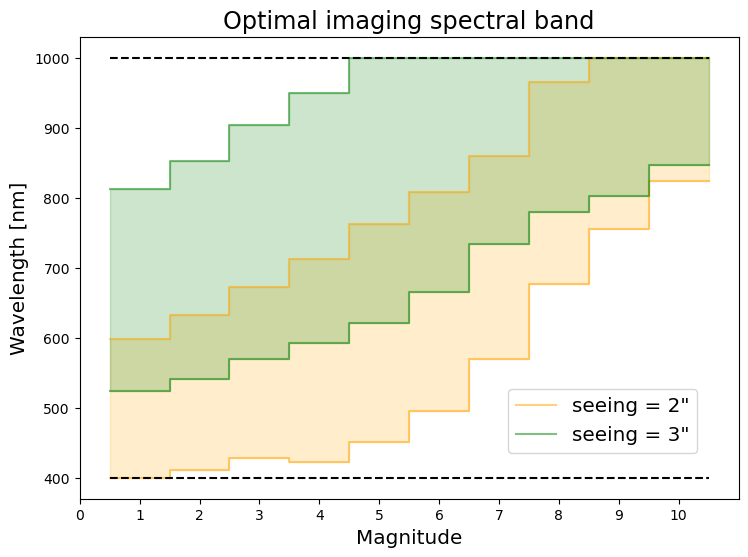}
    \caption{Optimal imaging band depending on the seeing and the object magnitude $M^v$. The vertical coloured bands give the optimal imaging band for the corresponding conditions, and the WFS uses the remaining flux. The black dashed lines indicate the imaging and WFS bandwidth boundaries.}
    \label{fig:optiband}
\end{figure}

In this subsection, we now jointly optimise the central wavelength and the bandwidth of the imaging band. In order to best examinate how the WFS and the imaging channel share photons, we consider here that the two channels share the same spectral band [400\,nm, 1000\,nm] (hence $\Delta\lambda_{wfs}=600$\,nm this time). For imaging bands larger than 50\,nm, the PSF is computed as the average of different PSFs at wavelengths separated by less than 50\,nm across the imaging band. We plot in Fig.~\ref{fig:optiband} the optimal imaging bands obtained for two different seeings, namely 2" (yellow curves) and 3" (green curves), as a function of the object's magnitude. For each combination of seeing and magnitude, the imaging band is given by the corresponding coloured vertical band, and the WFS uses the remaining flux.

What we can first note is that the trend remains consistent with the previous observations, \textit{i.e.}, the optimal imaging wavelengths get longer as the seeing or object magnitude increases. Regarding the optimal bandwidths, two regimes can be observed: a high flux regime, for $M^v\le5$, where the optimal bandwidth only slightly increases with the magnitude ; and a low flux regime, for $M^v>5$, where the optimal bandwidth decreases with the magnitude. In the latter case, the noise in the WFS measurements begins to limit the AO correction, hence the imaging channel must concede some photons to the WFS in order to maintain a certain level of correction. The optimal bandwidth in this regime is therefore the result of a trade-off between image SNR and quality of the AO correction. In the high flux regime, the WFS noise is not limiting and the imaging bandwidth has little (or no) influence on the AO correction level. Consequently, in this regime, the optimal imaging bandwidth results from a trade-off between the image SNR and the resolution of the average PSF. It is noteworthy to observe that, in the low flux regime, the optimal imaging bandwidth is greater for a seeing of 3" than 2". This is a consequence of the lower level of the OTF for larger values of the seeing, which in turn requires a better image SNR for deconvolution.

As compared to the results of Section~\ref{sec:lbd_ima}, we noted a reduction of several percent in RNMSE when the imaging bandwidth is optimised jointly with the imaging wavelength (w.r.t the case where the imaging bandwidth is fixed to 50\,nm). For object magnitudes $M^v\le4$, the gain is around 5\%, and it goes up to 10\% when $M^v=8$. The reduction is greater as the magnitude increases because the RNMSE value is more sensitive to the imaging bandwidth when the flux is low.  


\section{Discussion on the method's assumptions}\label{sec:disc_asp}
\subsection{Comparison to other metrics}\label{sec:ometrics}

\begin{figure}[ht!]
    \centering
    \includegraphics[scale=0.7]{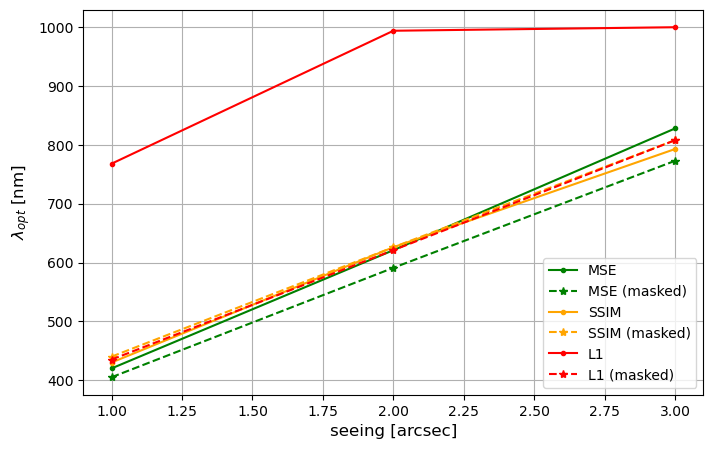}
    \caption{Optimal imaging wavelength as a function of the seeing for different metrics, with or without masking the background. The results are averaged over 20 noise realisations. $M^v$=4.}
    \label{fig:vs_metrics}
\end{figure}

Our decision to use the MSE as an image quality metric, and therefore as an optimisation criterion, is mainly motivated by the resulting simplification of calculations and the fact that it enables a frequency analysis of the error. However, we can ask ourselves whether or not the results we obtained would change significantly for other metrics. To address this question, we performed the same optimisations as in Sect.~\ref{sec:lbd_ima}, with the same parameters, but using other distances than the L2 norm, namely the L1 norm (or absolute difference) and the SSIM~\cite{wangImageQualityAssessment2004}. Since there is no analytical expression of the averaged error (Eq.~\eqref{eq:mse}) for these distances, we had to simulate focal plane images, deconvolve them and compute the average distance over several noise realisations and for the object shown in Fig.~\ref{fig:sats}. In Fig.~\ref{fig:vs_metrics}, we plot the optimal imaging wavelengths obtained with each metric as a function of the seeing for an object magnitude of 4 (see solid lines). It can be seen that the optimal imaging wavelengths obtained using the L2 norm and the SSIM are very close. This result is not especially surprising as it can be shown that these two distances are related~\cite{dosselmannComprehensiveAssessmentStructural2011, wangAssociationsMSESSIM2021}. The L1 norm however exhibits a completely different behaviour and leads to much longer optimal imaging wavelengths, for which the visual image quality (after deconvolution) is actually worse. We may wonder why there is such a difference between the L2 and L1 norms. The main difference between these two distances is that the L2 norm is dominated by pixels with the highest error amplitude, whereas the L1 norm gives all pixels the same weight in computation. The images we process consist of a bright compact object on a dark background, and the error between the reference and deconvolved images is greater where the intensity of the reference image is high. Consequently, the L2 norm is dominated by the error on the object, whereas the L1 norm is much more affected by the background, which comprises the majority of pixels. As a result, when using the L1 norm, the optimisation aims to reduce the error on the background, which is essentially low frequency. Since low frequencies are better reconstructed with a long imaging wavelength, the L1 norm causes the optimisation to converge towards longer wavelengths.

Considering the influence that the background may have on the behaviour of the distances, we also tested applying weightings when computing each distance. This is possible since the L2 norm, the L1 norm and the SSIM are all pixel-wise averaged values. In particular, we computed each distance using a binary mask centred on the object of interest in order to entirely discard the background from the distance computation (see dashed lines in Fig.~\ref{fig:vs_metrics}). We can see that using a mask has a minor impact on the results if we consider the L2 norm or the SSIM, implying that these distances where indeed dominated by the high amplitude errors on the object of interest. Regarding the L1 norm, the optimal imaging wavelengths obtained when using a mask are much shorter than when not, and they correspond to the ones obtained with the MSE and the SSIM. 

In conclusion, it appears that the L1 norm is not a judicious choice in our case as it is strongly influenced by the background if not computed on a mask. Regarding the SSIM and the MSE, they both gives similar results. As a result, using the MSE seems preferable as it is easier to compute and it enables a frequency analysis of the error.   


\subsection{Robustness to the noise model}\label{sec:vs_noise}

\begin{figure}[ht!]
    \centering
    \includegraphics[scale=0.7]{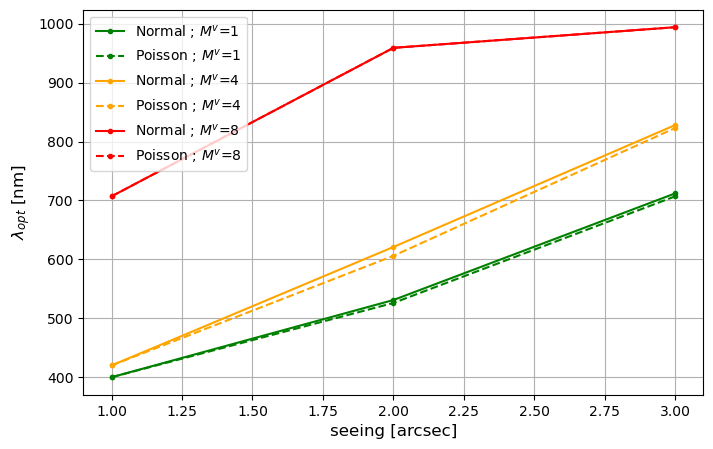}
    \caption{Optimal imaging wavelength in function of the seeing for different object magnitude and noise model, namely a normal homogeneous distribution (continuous lines) and a poisson distribution (dashed lines). The results are averaged over 20 noise realisations. For $M^v=8$, the curves are superimposed.}
    \label{fig:vs_noise}
\end{figure}

The analytical formula obtained for the MSE Eq.~\eqref{eq:mse_wiener} relies on the simplifying assumption that the noise is homogeneous. To assess the impact of this approximation on the results, we computed the error on average (just like in the previous section) using two different noise models. The first one is a Gaussian homogeneous distribution, as described in Sect.~\ref{sec:prior}. For the second model, we used a mixture of a Poisson distribution (for the photonic noise) and a Gaussian distribution (for the RON), which is assumed to accurately represent real noise. We plot in Fig.~\ref{fig:vs_noise} the optimal imaging wavelength as a function of the seeing for the two noise models considered and for different object magnitudes. It can be observed that, whether we are in a strong flux ($M^v$=1) or a weak flux ($M^v$=8) regime, the optimal imaging wavelengths obtained are very close for both noise models. The most significant difference occurs for a seeing of 2" and a magnitude of 4, but even then, the difference is of less than 25\,nm and both wavelengths will lead to a similar image quality. This confirms that the formula given in Eq.~\eqref{eq:mse_wiener} can be used to find the optimal imaging wavelength, even though it relies on an approximate noise model. 

\subsection{Robustness to PSF mis-estimation}

Another assumption used to obtain the analytical formula of Eq.~\eqref{eq:mse_wiener} is that the PSF is entirely known for the deconvolution. In practice, the PSF used for the deconvolution step, denoted $\mathbf{h_{dec}}$, has to be estimated from the image(s) using either multi-frame blind deconvolution algorithms (see, \textit{e.g.}, \cite{schulzMultiframeBlindDeconvolution1993, mugnierMyopicDeconvolutionWavefront2001} or \cite{asensioramosAcceleratingMultiframeBlind2023} for more recent works), or parametric model-based approaches as in~\cite{yanMarginalizedMyopicDeconvolution2023,Fetick-a-20}. 

\begin{figure}[ht!]
    \centering
    \includegraphics[scale=0.7]{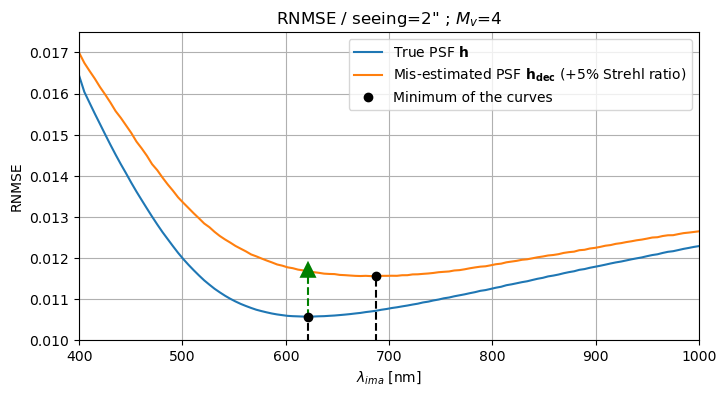}
    \caption{RNMSE as a function of the imaging wavelength, for seeing=2" and Mv=4. \textit{Blue}: RNMSE computed from the analytical formula of Eq.~\eqref{eq:mse_wiener}; \textit{Orange}: RNMSE computed from images deconvolved with a PSF $\mathbf{h_{dec}}$ having a Strehl ratio $\approx5$\% better than the true PSF $\mathbf{h}$. The green triangle indicates the value of the orange curve at the x-coordinate of the minimum of the blue curve.}
    \label{fig:mis_est}
\end{figure}

As a proxy to assess the influence of a mis-estimation of the PSF, we compute the average error obtained when deconvolving the focal plane images with a PSF, $\mathbf{h_{dec}}$, having a slightly better Strehl ratio (around 5\% better) than the true PSF, $\mathbf{h}$, used to simulate the images. In Fig.~\ref{fig:mis_est}, we plot this RNMSE as a function of the imaging wavelength (orange curve), as well as the RNMSE computed from Eq.~\eqref{eq:mse_wiener} (blue curve), which represents the average error when the deconvolution is performed using the true PSF $\mathbf{h}$. Both RNMSE curves are computed for a seeing of 2" and an object magnitude of 4. Let us denote by $\lambda_{opt}$ the optimal imaging wavelength when the PSF is known (\textit{i.e.}, the minimum of the blue curve), and by $\lambda_{mis}$ the optimal imaging wavelength when $\mathbf{h_{dec}}$ differs from $\mathbf{h}$ (\textit{i.e.}, the minimum of the orange curve). Figure \ref{fig:mis_est} shows that when the mis-estimation of the PSF is taken into account (orange curve), the optimal imaging wavelength $\lambda_{mis}$ shifts, here towards longer wavelengths, by a few tens of nanometers compared to $\lambda_{opt}$. However, since the RNMSE curves are quite flat near their minimum, using $\lambda_{opt}$ instead of $\lambda_{mis}$ does not increase the average error by much, as can be seen by comparing the y-coordinate of the green triangle with the minimum of the orange curve. In the example shown in Fig.~\ref{fig:mis_est}, the RNMSE difference is $1.3\times10^{-4}$, which corresponds to 1.12\% of the RNMSE value at the minimum of the orange curve. 

We repeated this process for different conditions (seeing and object magnitude), and obtained similar conclusions each time. In cases where the true PSF is mis-estimated (with less than 5\% of Strehl ratio error), using the optimal imaging wavelength $\lambda_{opt}$, given by Eq.~\eqref{eq:mse_wiener}, instead of the actual optimal wavelength $\lambda_{mis}$ only increases the RNMSE by less than 2\% of its value. 

In short, although the analytical criterion of Eq.~\eqref{eq:mse_wiener} gives an optimal imaging wavelength slightly different than if taking into account the mis-estimation of the PSF, the resulting loss in terms of RNMSE is very small. This analysis could be extended in the future, and an analysis involving a complete simulation and the choice of a deconvolution algorithm could be carried out.


\section{Toward a robust imaging system}\label{sec:ms}
\subsection{Problem statement}

The results presented in the previous sections suggest that the optimal imaging wavelength strongly depends on the turbulence conditions and object flux. Unfortunately, these are varying parameters. First, regarding the object flux, satellites may have very different average apparent magnitudes depending on their size and materials. Besides, the instantaneous apparent magnitude of a LEO satellite can vary during observation due to changes in its distance, its orientation relative to the Sun and Earth, and its proper rotation. Then, regarding turbulence conditions, it is well known that they can vary significantly during a single night of observation \cite{kellererAssessingTimeScales2007}. Additionally, the apparent seeing on the line of sight (LOS) is a function of the satellite's elevation since the lower the satellite is, the more atmosphere, and therefore turbulence, is traversed by the light~\cite{petitLEOSatelliteImaging2020}. Hence, when observing LEO satellites, whose elevation vary rapidly, the seeing also varies during the observation. Consequently, a system designed for satellite observation should be robust to seeing and magnitude variations, in the sense that it maintains acceptable performance for a wide range of conditions.

To address this issue, two solutions could be envisioned. The first one, hereafter labelled (A), is dynamically adjusting the imaging wavelength according to the instantaneous seeing and object magnitude. However, this solution is hard to implement in practice as it requires precisely adjusting the imaging wavelength in real time. Besides, having the imaging wavelength changing from one acquisition to another makes the exploitation of the images more complicated. Although this solution is hard to implement and complicates the data exploitation, we still consider it in the following as an ideal reference case for comparison with the other solutions: it always reaches optimal performance. The second solution we will focus on, hereafter labelled (B), is optimising the imaging wavelength on average over a range of seeing and magnitude values.


\subsection{On average co-design framework}

From a Bayesian point of view, optimising the imaging wavelength on average over a range of seeings and object magnitudes (solution (B)) corresponds to finding $\lambda^{opt}$ such that:
\begin{equation}
    \begin{split}
        \lambda^{opt} &= \argmin_{\lambda_{ima}} \left[ \mathbb{E}_{\textbf{o},\textbf{i},M^v,s}\left[\|\hat{\textbf{o}}(\lambda_{ima})-\textbf{o}\|^2\right] \right]\\
        &= \argmin_{\lambda_{ima}} \left[\int p(s)\int p(M^v) \mathbb{E}_{\textbf{o},\textbf{i}|M^v,s}\left[\|\hat{\textbf{o}}(\lambda_{ima})-\textbf{o}\|^2\right]\mathrm{d}M^v \mathrm{d}s \right],
    \end{split}
    \label{eq:lbd_bayes}
\end{equation}
where $p(M^v)$ is the prior probability of the object magnitude, $p(s)$ is the prior probability of the seeing $s$, and $\mathbb{E}_{\textbf{o},\textbf{i}|M^v,s}\left[\|\hat{\textbf{o}}-\textbf{o}\|^2\right]$ is the MSE for given conditions as defined in Eq.~\eqref{eq:mse_wiener}. The optimal imaging wavelength given by Eq.~\eqref{eq:lbd_bayes} is the one minimising the MSE on average under the prior law, \textit{i.e.}, on average over the conditions, and over noise and object outcomes. In order to obtain a robust system, \textit{i.e.}, one that maintains acceptable performance even in conditions that are  \textit{a priori} rare, we decide to consider a uniform prior law for the the seeing and for the object magnitude. This gives all the considered conditions the same weight in the calculation of the optimal imaging wavelength. Additionally, since we do not have an expression for the MSE as a function of the seeing and the object magnitude, we sample the seeing and object magnitude distributions, and we approximate the two integrals by sums. As a result, in practice, we compute the optimal imaging wavelength $\lambda^{opt}$ using the following formula:
\begin{equation}
    \begin{split}
        \lambda^{opt} &= \argmin_{\lambda_{ima}} \left[\overline{\mathrm{MSE}}(\lambda_{ima}) \right],\\
        \overline{\mathrm{MSE}}(\lambda_{ima}) &= \frac{1}{N_mN_s}\sum_{j=1}^{N_m}\sum_{l=1}^{N_s} \mathbb{E}_{\textbf{o},\textbf{i}|M^v_j,s_l}\left[\|\hat{\textbf{o}}(\lambda_{ima})-\textbf{o}\|^2\right] ,
    \end{split}
    \label{eq:opt_avg}
\end{equation}
where $\{M^v_j\}_{1\le j\le N_m}$ and $\{s_l\}_{1\le l\le N_s}$ are respectively the $N_m$ object magnitude and $N_s$ seeing samples. To sum up, the optimal imaging wavelength $\lambda^{opt}$, computed on average over the conditions $\{M^v_j, s_l \}_{j,l}$, is obtained by minimising an average error $\overline{\mathrm{MSE}}$ which is simply the average of the different MSE curves obtained for each considered conditions (just like the ones of Fig.~\ref{fig:lbd_ima}).\newline 

Let us now compare the two proposed solutions, namely: (A) adapting the imaging wavelength in real time according to the seeing and the object magnitude; (B) optimising the imaging wavelength on average over the conditions using Eq.~\eqref{eq:opt_avg}. We consider two systems, (A) and (B), for which the imaging wavelength is chosen according to their respective solutions. For both systems, the imaging bandwidth is fixed to 200\,nm, which is consistent with the results presented in Fig.~\ref{fig:optiband}. As in Sect.~\ref{sec:lbd_ima}, the imaging wavelengths of the two systems must lie within the range [400\,nm; 1000\,nm]. The imaging wavelength of system (B) is optimised on average over the seeings $s\in\{1",2",3",4"\}$ and the object magnitudes $M^v\in\{0,2,4,6,8\}$, making a total of 20 condition pairs. After computing the average error $\overline{\mathrm{MSE}}$, we found an optimal imaging wavelength of 838\,nm for system (B). 

\begin{figure}[htb!]
    \centering
    \includegraphics[scale=0.85]{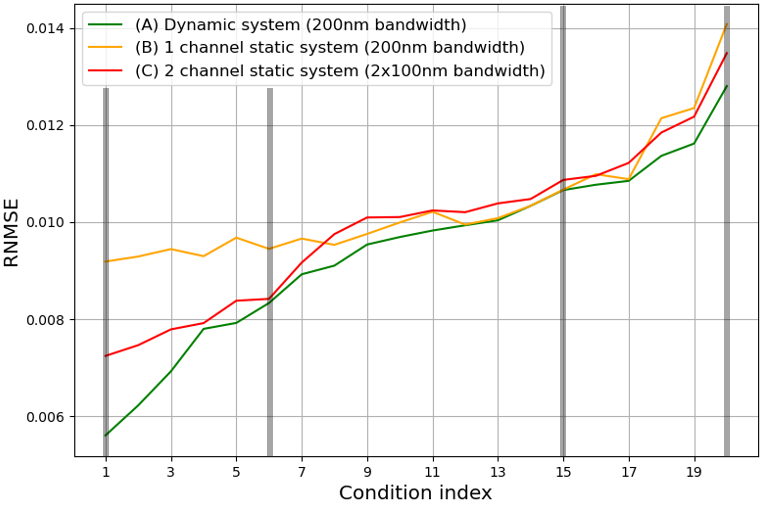}
    \caption{RNMSE of three different systems (A, B and C) as a function of the conditions (seeing and object magnitude) ordered from the best conditions (on the left) to the worst (on the right), according to the RNMSE of system (A). The vertical grey lines correspond to the four pairs of conditions that are illustrated in Fig.~\ref{fig:sat_ms}, in the same order.}
    \label{fig:comp_syst}
\end{figure}

Figure~\ref{fig:comp_syst} shows the RNMSE computed for the two systems as a function of the conditions, ordered from the best (on the left), to worst conditions (on the right). The 20 pairs of conditions are ordered according to the RNMSE of the reference system (A). Indeed, since this system is always optimal, its RNMSE increases when the conditions deteriorate. As it can be observed on Fig.~\ref{fig:comp_syst}, from the $12^\mathrm{th}$ to the $17^\mathrm{th}$ condition pair, the dynamic system (A) (green curve) and the static system (B) (yellow curve) have very similar RNMSE values. This corresponds to conditions for which the imaging wavelength of the dynamic system (A) is close to that of system (B) (838\,nm). For condition indexes higher than 17, the dynamic system (A) performs slightly better than the static system (B). In this regime, system (B) is limited by the residual phase error which increases as the conditions get worse. The dynamic system (A) also suffers from the increase of the residual phase error, but its imaging wavelength increases (up to 1000\,nm for the $20^\mathrm{th}$ condition pair) to attenuate this effect. For favourable conditions, \emph{i.e.}, for condition indexes below 12, the dynamic system (A) performs better than the static system (B) as well, and the difference in RNMSE increases noticeably as the conditions improve. Here, system (B) is primarily limited by diffraction, with a resolution capped at the diffraction limit. Hence, for indexes lower than 12, the RNMSE of system (B) doesn't really improve with improving conditions. The slight decrease in RNMSE observed when the conditions improve is due to an improvement in image SNR, which enables more effective noise removal. Regarding the dynamic system (A), its imaging wavelength decreases when the conditions get better (down to 400\,nm for the first condition pair), enablinegg the obtention of sharper images and a reduction of the RNMSE.

By its definition, system (B) is the system that minimises the MSE on average over all the considered conditions, under the constraint of using a single static imaging channel of 200\,nm bandwidth. Hence, system (B) is simpler than system (A) and robust (to variations of conditions), but at the cost of a reduced performance under favourable conditions. In summary, the framework proposed in this subsection helps finding a robust system under simplicity design constraints, which may be very useful in practice. If we now want to improve performance of system (B), we must adjust the design constraints and accept an increased complexity.


\subsection{On average, multi-channel co-design framework}\label{sec:mc_framework}

To mitigate the limitations of solution (B), we now consider a third solution, hereafter labelled (C), which is using simultaneously several imaging channels at different imaging wavelengths, jointly optimised on average over a range of conditions. In the case of a multi-channel acquisition system with $N_c$ channels, the framework presented in Sect.~\ref{sec:model} must be slightly extended to take the different acquisition channels into account. 

Firstly, regarding the imaging model, nothing changes except that $N_c$ images $\{\textbf{i}_k\}_{1\le k\le N_c}$ are recorded instead of just one, with:
\begin{equation}
    \forall k\in\llbracket1,N_c\rrbracket: \textbf{i}_k\ =\ N_{ph,k}\left[\textbf{h}_k\ast \textbf{o}\right] + \textbf{n}_k,
    \label{eq:mc_imaging}
\end{equation}
where $\textbf{i}_k$, $\textbf{h}_h$ and $\textbf{n}_k$ are respectively the image, the PSF and the noise in the $k^\mathrm{th}$ channel. The number of photoelectrons in each channel $N_{ph,k}$, which depends on the solar spectrum and the object albedo, is explicitly included in Eq.~\eqref{eq:mc_imaging} such that here $\sum \textbf{o}=1$. Hence, we set $A=1$ for the object prior PSD $S_o$ of Eq.~\eqref{eq:psdo}. Note that Eq.~\eqref{eq:mc_imaging} implicitly assumes that the object is gray, \textit{i.e.}, that it is the same in all channels, except for its total flux. This is an approximation as the different materials composing the satellite may have distinct spectral signatures. This approximation is legitimate as we do not have prior spectral information on the object for the system design. The different noise terms $\{\textbf{n}_k\}$ are assumed to be mutually independent and remain distributed according to an homogeneous Gaussian distribution with zero mean. Assuming that the RON and the number of pixels $N_{pix}$ are the same in all channels, the noise PSDs $\{S_{n,k}\}$ are given by $S_{n,k}=N_{ph,k} + \sigma^2_{ron}\times N_{pix}$. 

These $N_c$ images are then jointly processed to produce a single estimate $\hat{\textbf{o}}(\{\textbf{i}_k\})$. In this case, the MMSE estimator remains linear and corresponds to a multi-channel Wiener filter:
\begin{equation}
    \tilde{\hat{\mathrm{o}}}_{mmse}(f_j) = \frac{\sum_{k=1}^{N_c} \left[N_{ph,k}S_{n,k}^{-1}(f_j)\tilde{\mathrm{h}}_k^*(f_j)\tilde{\mathrm{i}}_k(f_j)\right] + S_o^{-1}(f_j){\tilde{\overline{\mathrm{o}}}}(f_j)}{\sum_{k=1}^{N_c}\left[ N_{ph,k}^2S_{n,k}^{-1}(f_j)\vert\tilde{\mathrm{h}}_k(f_j)\vert^2\right]+S_o^{-1}(f_j)}.
    \label{eq:mc_wiener}
\end{equation}
The resulting MSE, knowing the conditions, is given by:
\begin{equation}
    \text{MSE}(\hat{\textbf{o}}_{mmse})\ =\ \frac{1}{N_{pix}}\sum_{j=0}^{N_{pix}-1} \frac{S_o(f_j)}{\sum_{k=1}^{N_c} \left[ N_{ph,k}^2\vert\tilde{\mathrm{h}}_k(f_j)\vert^2 \frac{S_o(f_j)}{S_{n,k}(f_j)}\right]+1},
    \label{eq:mc_mse}
\end{equation}
and the RNMSE is equal to $\sqrt{\mathrm{MSE}}$ (as the object is already normalised here). Note that, for a single channel, this corresponds to the formulas of Eq.~\eqref{eq:mse_wiener} and Eq.~\eqref{eq:rnmse}, hence Eq.~\eqref{eq:mc_mse} is a generalization of these formulas. Finally, the $N_c$ optimal imaging wavelengths optimised on average over the conditions can be computed using the formula of Eq.~\eqref{eq:opt_avg}, except that the MSE is now the one of Eq.~\eqref{eq:mc_mse} and that $\lambda^{opt}$ is replaced by a set of optimal wavelengths $\{\lambda_k^{opt}\}_{1\le k\le N_c}$. 

\begin{figure}[h!]
    \centering
    \includegraphics[scale=0.5]{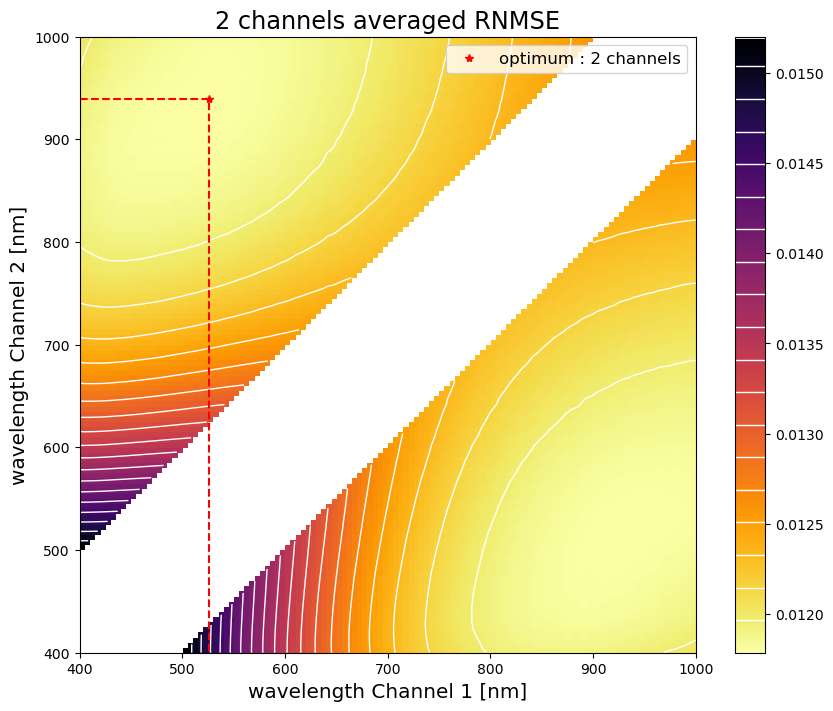}
    \caption{Root of the MSE map, averaged over seeings ranging from 1" to 4" and object magnitudes ranging from 0 to 8, as a function of the two channels imaging wavelength \{$\lambda_1;\lambda_2$\} ($\Delta\lambda=100$\,nm).}
    \label{fig:ms_err}
\end{figure}

In order to compare this third solution (C) to the two first solutions (A) and (B) proposed above, we consider a third system (C) composed of two imaging channels of 100\,nm bandwidth each. The wavelengths $\{\lambda_1,\lambda_2\}$ of the two imaging channels of system (C) are optimised on average over the same conditions as for system (B), \textit{i.e.}, over seeings $s\in\{1",2",3",4"\}$ and object magnitudes $M^v\in\{0,2,4,6,8\}$. In Fig.~\ref{fig:ms_err}, we plot the square root of the average error $\overline{\mathrm{MSE}}$ as a function of the imaging wavelengths \{$\lambda_1;\lambda_2$\}.
The white region corresponds to couples of imaging wavelengths \{$\lambda_1;\lambda_2$\} that are impossible due to overlap of the imaging bands. The figure is symmetrical about the first bisector as we consider that the two imaging channels are identical, except for their wavelength. The minimum of the error map, indicated by the red star, is reached for $\lambda_1=526$\,nm and $\lambda_2=939$\,nm, \textit{i.e.}, a short and a long wavelength. The short wavelength $\lambda_1$ is mostly used when the seeing and the magnitude are small, and \textit{vice versa} for $\lambda_2$. More precisely, according to the results of Sect.~\ref{sec:rmse_spectrum}, the short wavelength helps best restoring the mid and high spatial frequencies while the long wavelength helps restoring the low frequencies. It is worth mentioning that when the optimisation is performed for given conditions (\textit{i.e.}, when computing the MSE of Eq.~\eqref{eq:mc_mse}), it always gives two optimal imaging wavelengths separated by 100\,nm. In other words, for given conditions, the optimisation tends to recreate a single-channel imaging system whose central imaging wavelength ($(\lambda_1+\lambda_2)/2$) is the optimal one (for the bandwidth and conditions that are considered).

Now, regarding the performance of system (C), we can observe on Fig.~\ref{fig:comp_syst} that its RNMSE (red curve) is close to that of the dynamic reference system (A) (green curve) for all the conditions. Or, put another way, as compared to the single-channel static system (B), system (C) is a robust system in the sense that it keeps quasi-optimal performance in all conditions.

\subsection{Comparison of the different solutions: illustrations}

\begin{figure}[h!]
    \centering
    \includegraphics[width=\textwidth]{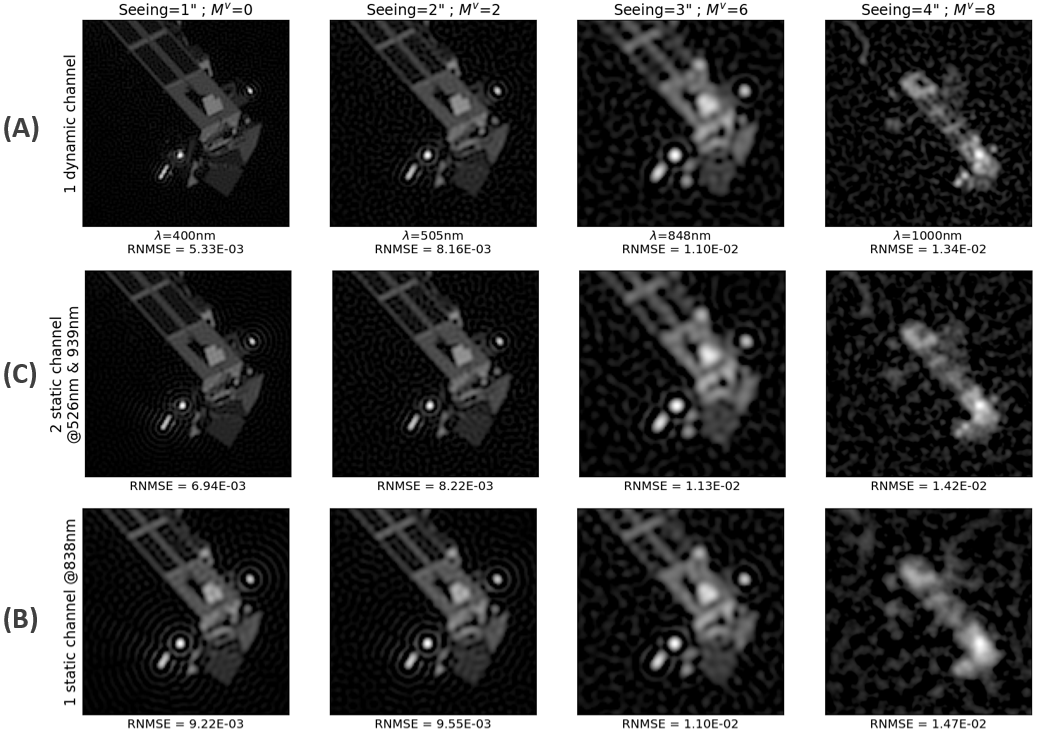}
    \caption{Simulated images, after deconvolution, of observations of the satellite Envisat made by the three systems: A (top row), C (middle row) and B (bottom row); and for different conditions of observation. The RMSE scores indicated under each image are computed from the difference with the reference image.}
    \label{fig:sat_ms}
\end{figure}

To conclude this section, we now present some illustrations showing the performance of the three proposed solutions, namely: (A) adapting the imaging wavelength in real time (reference system); (B) optimising the imaging wavelength on average over the conditions (single-channel); and (C) optimising the imaging wavelengths of a multi-channel imaging system on average over the conditions as well. We plot in Fig.~\ref{fig:sat_ms} some restored images of simulated Envisat observations made by each system for different combinations of seeing and object magnitude.
As in Fig.~\ref{fig:comp_syst}, the best conditions are displayed on the left and the worst on the right. The images displayed in Fig.~\ref{fig:sat_ms} are consistent with the RNMSE values of the three systems plotted in Fig.~\ref{fig:comp_syst}. First of, we can see that, in all conditions, the dynamic reference system (A) provides the best visual image quality among the three systems. We can also observe that the visual quality of the images obtained by the two-channel system (C) is only slightly worse than that of the reference system (A), for all the conditions as well. At the opposite, the single-channel system (B) does not match the reference system (A) in terms of visual image quality for all conditions. In particular, under good conditions, it is clear that the resolution of system (B) is limited by diffraction and that the visual image quality does not change with the conditions (see the first two columns in the last row in Fig.~\ref{fig:sat_ms}). Additionally, in conditions where the single-channel system (B) is optimal, and therefore better than system (C) (see, for example, the third column of Fig.~\ref{fig:sat_ms}), the difference in visual image quality between systems (B) and (C) is negligible. As a result, we can confirm the conclusions previously established in the two last sub-sections, namely: \textit{i)} Using only one imaging wavelength, optimised on average, is not sufficient to cover a large range of conditions; and \textit{ii)} the proposed multi-channel solution (C) is an effective and easy-to-implement alternative to the reference dynamic solution (A).


\section{Conclusion}\label{sec:ccl}

In this article, we applied a co-design approach to the design of AO-assisted imaging systems. This approach is based on the minimisation of a MSE-based criterion, which takes into account the effects of the entire optical/digital system, as well as the object properties, and which is easy to compute. To illustrate the interest of this approach, we studied the optimisation of the wavelength distribution between the imaging and WFS channels for satellite imaging. We observed that the optimal imaging wavelengths were getting longer as the seeing increases, which is consistent with the literature, or as the magnitude increases, a dependency that is not possible to observe using traditional PSF-related criteria. Besides, the optimal imaging wavelengths obtained when using the MSE-based method lead to better apparent image quality after deconvolution than the wavelengths found using PSF-based criteria. 

Since the seeing and the object magnitude are variables which can vary from an observation to another and during the observation itself, we explored solutions to make a system robust to seeing and magnitude fluctuations, namely, we considered optimising the imaging wavelength on average over a range of seeings and object magnitudes. This approach can help making a system robust, but in the limits of the constraints applied to the system design. For instance, in this case, we noticed that using a unique imaging wavelength was not sufficient to reach quasi-optimal performance over wide range of conditions. Hence, we then considered using several imaging channels at different wavelengths and we extended our MSE-based optimisation framework accordingly. In particular, we showed that a two-channel system reaches quasi-optimal performance over a wide range of external conditions. This could be further improved, for example by considering a third imaging channel or by jointly optimising the central wavelength and bandwidth of each imaging channel. Beyond the case study that has been chosen here, namely satellite observation, the results presented in this article remain valid for the observation of other satellite-like objects, such as asteroids.   

This work has many perspectives. Firstly regarding the flux distribution issue, we used---for simplicity---rectangular functions to describe the spectral bands of the different system channels, but more complex distributions could be considered. Secondly, we discussed the impact of not knowing perfectly the PSF, and it would be interesting to further study this problem by a complete simulation with a myopic deconvolution. Then, in this article we focused on the optimisation of the imaging and WFS bandwidth only, which includes a trade-off between WFS noise error and image SNR, with the other system parameters fixed. Other parameters of the AO-system, such as the AO loop frequency or the number of actuators of the deformable mirror, could be additional degrees of freedom of the optimisation. Jointly optimising all the system parameters would result in a trade-off between all the AO error budget terms (\textit{e.g.}, temporal error, fitting, etc.), and while these terms are generally equally distributed in AO designs, using the MSE criterion might lead to an optimum where the different error budget terms are unevenly distributed. Finally, it would be interesting to verify these results, eventually, with an experimental setup.

\begin{backmatter}
\bmsection{Funding} Office National d’études et de Recherches Aérospatiales; Agence Nationale de la Recherche (ANR-22-EXOR-0016).

\bmsection{Acknowledgment} The authors thank Ugo Tricoli for providing the satellite images used in this article, and Romain F\'etick for our interactions and his help on the PSF simulation codes. They also thank the reviewers for their careful reading and valuable suggestions. We acknowledge support from project PEPR Origins, reference ANR-22-EXOR-0016, supported by the France 2030 plan managed by Agence Nationale de la Recherche.

\bmsection{Disclosures} The authors declare no conflicts of interest.
\end{backmatter}

\bibliography{Acronymes,EnglishAcronyms,biblio}

\end{document}